\documentclass[a4paper]{jpconf}
\usepackage{graphicx}
\usepackage{amsmath}
\usepackage{amsfonts}
\usepackage{amssymb}
\begin{document}
\title[Tsallis]{Large Transverse Momenta and Tsallis Thermodynamics}

\author{J.~Cleymans}
\address{UCT-CERN Research Centre and Physics Department, University of Cape Town, South Africa}
\ead{jean.cleymans@uct.ac.za}
\author{M.~D.~Azmi}
\address{HEP Lab, Physics Department, Aligarh Muslim University, Aligarh - 202002, India}
\ead{danish.hep@gmail.com}   

\begin{abstract}
The charged particle transverse momentum ($p_T$) spectra measured by the ATLAS and CMS collaborations in proton - proton collisions 
at $\sqrt{s}$ = 0.9 and 7 TeV have been studied using Tsallis thermodynamics. A thermodynamically consistent form of the 
Tsallis distribution is used for fitting the transverse momentum spectra at mid-rapidity. It is found that the fits based on 
the proposed distribution provide an excellent description over 14 orders of magnitude with $p_T$ values up to 200 GeV/c.
\end{abstract}

\section{Introduction}
It is by now   well-known that the Tsallis distribution gives excellent fits to the transverse momentum distributions observed at
the Relativistic Heavy Ion Collider (RHIC)~\cite{STAR,PHENIX1,PHENIX2} and at the Large Hadron 
Collider (LHC)~\cite{ALICE,CMS1,CMS2,ATLAS,ALICE2} with only three parameters, $q$, $T$ and $dN/dy$ (or, alternatively, a 
volume $V$~\cite{ryb,cleymans1,azmi-cleymans}). The parameter $q$ is referred  to as the Tsallis parameter and 
discussed elsewhere~\cite{wilk4} in detail. 

It was recently shown that these fits extend to values~\cite{CMS3} of $p_T$ up to 200 GeV/c~\cite{wilk2,tsallis-poland}. 
This is unexpected because in this kinematic range hard scattering processes become important~\cite{wilk3}. 
A description of the high $p_T$ results has been discussed in \cite{Urmossy} where a model using a combination of Tsallis at 
low $p_T$ and QCD hard scattering at high $p_T$ was considered. 
The present analysis shows that the Tsallis distribution describes measurements up to the highest $p_T$ using the same 
parameters as those obtained at low $p_T$.

A power law based on the Tsallis distribution~\cite{tsallis1} is used to fit the $p_T$ spectra of charged particles measured by the 
ATLAS and CMS collaborations.  The ATLAS collaboration has reported the transverse momentum in an inclusive phase space 
region taking into account at least two charged particles in the kinematic range $|\eta| < 2.5$ and $p_T > 100$ MeV~\cite{ATLAS}. 
The CMS collaboration has presented the differential transverse momentum distribution covering a $p_T$ range up to 200 GeV/c, the 
largest range ever measured in a colliding beam experiment~\cite{CMS3}. 

The results can be compared to those obtained in~\cite{wilk2,tsallis-poland,wilk3,wilk5,wilk6} where very good fits to 
transverse momentum distributions were presented. We confirm the quality of the fits but obtain  
different values of the  parameters albeit using a different version of the Tsallis model.

\section{Tsallis Distribution}
The Tsallis distribution is defined as
\begin{equation}
f(E) \equiv \left[ 1 + (q-1) \frac{E-\mu}{T}\right]^{-\frac{1}{q-1}} ,
\end{equation}
which at large energies  behaves as
\begin{equation}
\lim_{E\rightarrow\infty} f(E) = \left(\frac{E}{T}\right)^{-\frac{1}{q-1}}  ,
\end{equation}
so that the scale is set by $T$ and the asymptotic behaviour is set by $q$. 
For high energy physics a consistent form  of Tsallis 
thermodynamics for 
the particle number density $n$, energy density, $\epsilon$, and pressure $P$ are given by~\cite{worku1,worku2} 
\begin{eqnarray}
n &=& g\int\frac{d^3p}{(2\pi)^3}
\left[ 1 + (q-1) \frac{E-\mu}{T}\right]^{-\frac{q}{q-1}} ,\\
\epsilon &=& g\int\frac{d^3p}{(2\pi)^3}E
\left[ 1 + (q-1) \frac{E-\mu}{T}\right]^{-\frac{q}{q-1}} ,\\
P &=& g\int\frac{d^3p}{(2\pi)^3}\frac{p^2}{3E}
\left[ 1 + (q-1) \frac{E-\mu}{T}\right]^{-\frac{q}{q-1}} .
\end{eqnarray}
where $T$ and $\mu$ are the temperature and the chemical potential,
$V$ is the volume and  $g$ is the degeneracy factor.
This introduces only one new parameter $q$ which 
for transverse momentum spectra is  always close to 1.
The  consistency conditions 
\begin{equation}
d\epsilon = Tds + \mu dn,~~~~~~~~~~~~~~~~~dP = nd\mu + s dT,
\end{equation}
are  satisfied. This is shown explicitly for one of  the relation $n = {\partial P}/{\partial \mu}$.
\begin{eqnarray*}
\frac{\partial P}{\partial\mu} &=& g\int\frac{d^3p}{(2\pi^3)} \frac{p^2}{3E} \frac{\partial}{\partial\mu} f^q\nonumber \\
&=& -g\int\frac{d^3p}{(2\pi^3)} \frac{p^2}{3E} \frac{d}{dE} f^q\nonumber \\
&=& -g\frac{4\pi}{(2\pi^3)} \int_0^\infty dp~\frac{p^4}{3E} \frac{d}{dE} f^q\nonumber \\
&=& -g\frac{4\pi}{(2\pi^3)} \int_0^\infty dp~\frac{p^3}{3} \frac{d}{dp} f^q~~~~~\text{using}~~~~EdE = pdp\nonumber \\
&=& g\frac{4\pi}{(2\pi^3)} \int_0^\infty dp~p^2 f^q\nonumber \\
&=& n\nonumber 
\end{eqnarray*}
\section{Transverse Momentum Distributions}
From the expression for the total number of particles: 
\begin{equation}
N = gV \int\frac{d^3p}{(2\pi)^3}
\left[ 1+(q-1)\frac{E-\mu}{T} \right]^{-\frac{q}{q-1}}    ,
\end{equation}
we obtain the corresponding momentum distribution 
\begin{equation}
E\frac{dN}{d^3p} = gVE\frac{1}{(2\pi)^3}
\left[1+(q-1)\frac{E-\mu}{T}\right]^{-\frac{q}{q-1}} .
\end{equation}
In terms of the rapidity and transverse mass variables, 
$E = m_T\cosh y$, 
this becomes
(at mid-rapidity $y=0$ and  for $\mu$ = 0)
\begin{equation}
\left.\frac{d^2N}{dp_T~dy}\right|_{y=0} 
= gV\frac{p_Tm_T}{(2\pi)^2}
\left[ 1+(q-1)\frac{m_T}{T} \right]^{-\frac{q}{q-1}},
\end{equation}
and, for charged particles it is given by a sum over the most abundant ones, $\pi^\pm, K^\pm, p, \bar{p}$.
The results are shown in figures (1) and (2) and have been discussed in more detail in~\cite{epjc}.
\begin{equation}
\left.\frac{d^2N(\text{charged})}{dp_T~dy}\right|_{y=0} 
= \sum_{i=\pi,K,p,...} g_iV\frac{p_Tm_T}{(2\pi)^2}
\left[ 1+(q-1)\frac{m_T}{T} \right]^{-\frac{q}{q-1}},
\end{equation}
The resulting parameters are listed in table 1.
%
\begin{figure}[h]
\begin{minipage}{17pc}
\includegraphics[width=17pc]{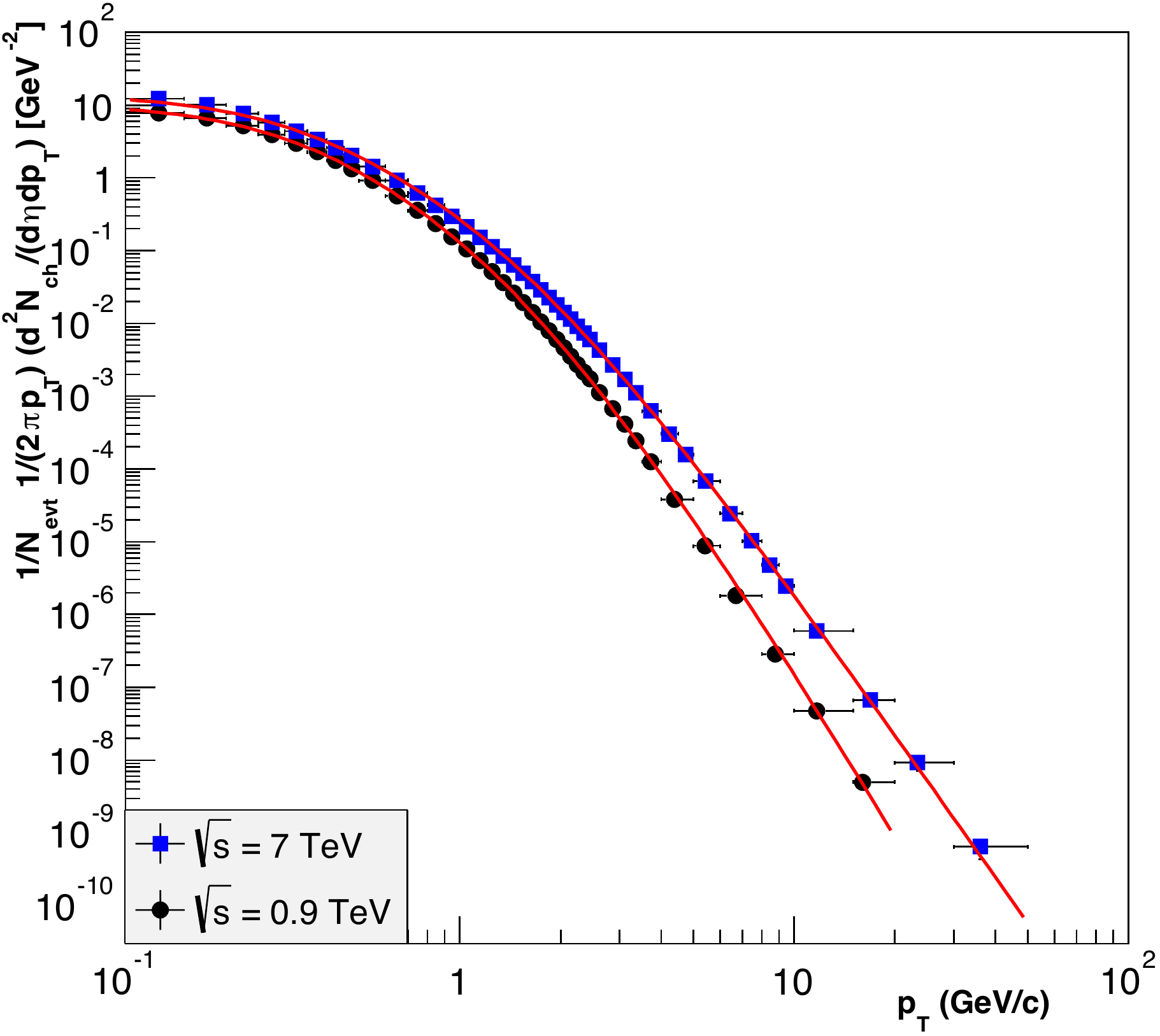}
\caption{Charged particle multiplicities as a function of the transverse momentum measured by the ATLAS detector 
for events with $n_{ch} \ge 2$, $p_T > $100 MeV and $|\eta| < $ 2.5 at $\sqrt{s}$ = 0.9 and 7 TeV in 
proton - proton collisions~\cite{ATLAS} fitted with Tsallis distribution.}
\end{minipage}\hspace{2pc}%
\begin{minipage}{17pc}
\includegraphics[width=17pc]{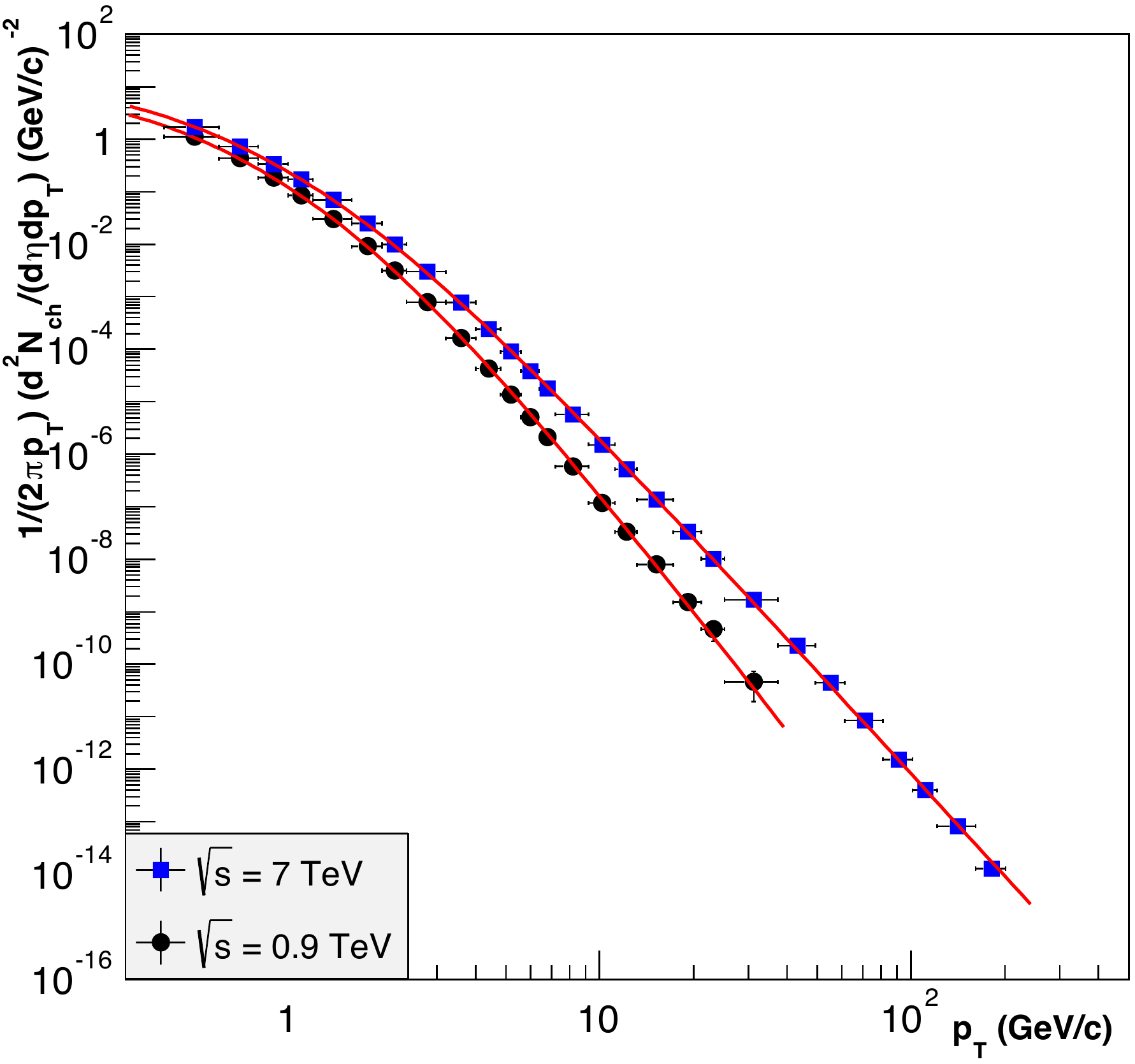}
\caption{Charged particle differential transverse momentum yields measured within $|\eta| < 2.4$ by the CMS detector in proton - proton collisions at $\sqrt{s}$ = 0.9 and 7 TeV~\cite{CMS3} fitted with Tsallis distribution.}
\end{minipage} 
\end{figure}

\begin{table}
\caption{Values of the $q$, $T$ and $R$ parameters and $\chi^2/NDF$ obtained from fits to the $p_T$ spectra measured by the ATLAS~~\cite{ATLAS} and CMS~\cite{CMS3} detectors.}
\label{par1}
\begin{tabular}{lccccc}
\hline
{\bf Experiment} & \multicolumn{1}{c}{\bf $\sqrt{s}$ (TeV)} & \multicolumn{1}{c}{\bf $q$} & \multicolumn{1}{c}{\bf $T$ (MeV)} & \multicolumn{1}{c}{\bf $R$ (fm)} & \multicolumn{1}{c}{\bf $\chi^2/NDF$} \\
\\
\hline
ATLAS & 0.9 & 1.129 $\pm$ 0.005 & 74.21 $\pm$ 3.55 & 4.62 $\pm$ 0.29 & 0.657503/36 \\
ATLAS & 7 & 1.150 $\pm$ 0.002 & 75.00 $\pm$ 3.21 & 5.05 $\pm$ 0.07 & 4.35145/41 \\
\hline
CMS & 0.9 & 1.129 $\pm$ 0.003 & 76.00 $\pm$ 0.17 & 4.32 $\pm$ 0.29 & 0.648806/17 \\
CMS & 7 & 1.153 $\pm$ 0.002 & 73.00 $\pm$ 1.42 & 5.04 $\pm$ 0.27 & 0.521746/24 \\
\hline
\end{tabular}
\end{table}

\section{Conclusion}
It is quite remarkable that the  transverse momentum distributions measured up to 200 GeV/c in $p_T$ can be described 
consistently over 14 orders of magnitude by a straightforward Tsallis distribution. 
The advantages are that the  thermodynamic conistency conditions are satisfied:
$$n = \frac{\partial P}{\partial \mu}~~~~~\text{etc...} ,$$
and the parameter $T$ truly deserves its name since 
$ T = {\partial E}/{\partial S}$.

\end{document}